\documentclass[useAMS,usenatbib]{mnras}
\usepackage{amsmath}
\usepackage{graphicx}
\usepackage{txfonts}
\usepackage{natbib}
\usepackage{bm}
\usepackage{changes}
\usepackage{multirow}
\usepackage{caption}
\usepackage{nicematrix}
\usepackage{enumitem}
\usepackage{xcolor}
\usepackage{tcolorbox}

\def\beq#1{\begin{equation}\label{#1}}
\def\eeq{\end{equation}}
\def\beqa#1{\begin{eqnarray*}\label{#1}}
\def\eeqa{\end{eqnarray*}}

\def\myfrac#1#2{\left(\frac{#1}{#2}\right)}

\def\comment#1{\relax}

\tcbset{width=0.5\textwidth,boxrule=0pt,colback=red!50,arc=0pt,left=0pt,right=0pt,boxsep=0pt}

\newcommand{\V}{V444 Cyg}
\newcommand{\kms}{km~s$^{-1}$}
\newcommand{\syr}{s~yr$^{-1}$}

\begin{document}

\title{Spectroscopic Searches for Evolutionary Orbital Period Changes in WR+OB Binaries: the case of V444 Cyg}
\author[A. Cherepashchuk et al.] {I. Shaposhnikov$^{1,2}$
\thanks{E-mail: iv.shaposhnikov@gmail.com, cherepashchuk@gmail.com, kpostnov@gmail.com},
A. Cherepashchuk$^1$, A. Dodin$^1$,  K. Postnov$^{1,3}$\\
$^{1}$Moscow State University, Sternberg Astronomical Institute, Universitetskij pr. 13, 119234 Moscow, Russia\\
$^{2}$Moscow State University, Faculty of Physics, Leninskiye Gory 1-2, 119991 Moscow, Russia\\
$^{3}$Kazan Federal University, Kremlyovskaya 18, 420008 Kazan, Russia\\
}	

\maketitle
\date{Received ... Accepted ...}
\pagerange{\pageref{firstpage}-\pageref{lastpage}} \pubyear{2023}

\label{firstpage}

\begin{abstract}
We present the results of new photometric and spectroscopic observations of WN5+O6 binary V444 Cyg and a detailed analysis of extant spectroscopy and photometry. Using elements of the spectroscopic orbit and assuming $e \approx 0$, $i \approx 78^{\circ}$ we determined the masses and orbit sizes of the components of V444 Cyg $M_{\mathrm{O6}} \approx 26.4~M_{\sun}$, $M_{\mathrm{WN5}} \approx 10.7~M_{\sun}$,  $a_{\mathrm{O6}} \approx 10.6~R_{\sun}$, $a_{\mathrm{WN5}} \approx 26.1~R_{\sun}$. Based on new and archival light curves and by applying the Hertzsprung method, we improved the photometrical estimate of secular increase rate of the orbital period in V444 Cyg $\dot{P}_{\mathrm{ph}} = 0.119\pm0.003$ s~yr$^{-1}$. From the comparison of the new and archival radial velocity curves of V444 Cyg we independently derived the secular orbital period change rate $\dot{P}_{\mathrm{sp}} = 0.147\pm0.032$ s~yr$^{-1}$, in agreement with the photometrical $\dot{P}_{\mathrm{ph}}$.  The obtained secular increase rate of the binary orbital period $\dot{P}$ and the mean radii of the components enabled us to estimate the stellar wind mass-loss rate from WR star $\dot{M}_{\mathrm{WN5}} = -(6.0\pm 0.4)\cdot 10^{-6}~M_{\sun}\mbox{ yr}^{-1}$.
\end{abstract}


\section{Introduction}
\defcitealias{Shaposhnikov2023}{Paper~I}

Wolf-Rayet (WR) stars appear at late evolutionary stages of massive stars preceding core-collapse supernova explosions and the formation of compact objects. 
Their most prominent spectral features include intense and broad emission lines of strongly ionized CNO elements which are formed in a powerful high-velocity stellar wind  
($\dot{M}_{\mathrm{WR}} \approx 10^{-5}~M_{\sun}\mbox{ yr}^{-1}$, $v \approx 10^3$ km s$^{-1}$).
Along with the mass, radius and chemical composition, the stellar wind mass-loss rate is one of the key parameters governing the evolution of a WR star. 

In addition to model-dependent methods of  analysis of emission line profiles in WR spectra \citep{Lamers1999}, observations of short-period close binaries with WR stars enable an independent estimation of $\dot{M}_{\mathrm{WR}}$

WR+OB systems are the most widespread among this class of close binaries. 
The stellar wind mass-loss from the components can be estimated from measurements of the secular change rate of the binary orbital period $\dot{P}$ due to the balance of angular momentum. Until recently, the secular change of the binary orbital period for several WR+OB systems has been inferred from the comparison of light curves taken at different epochs. However, in \cite{Shaposhnikov2023} (\citetalias{Shaposhnikov2023} below), we have justified the use of radial velocity curves taken at different epochs to search for secular orbital period changes in these close binaries. 

In the present paper, we confirm the applicability of this approach using new optical photometric and low-resolution spectroscopic observations of V444 Cyg = WR 139 = HD 193576 = HIP 100214 (WN5+O6, $B=8.53^m$, $V=8.00^m$).  
This is apparently the best studied WR+OB system. Its binary period change was discovered photometrically in 1974 \citep{Khaliullin1974} and has been confirmed by many later studies. 

In Section 2 we review basic photometric and spectroscopic information on V444 Cyg. In Section 3 we present our new observations. In Section 4 we search for changes in the orbital period using both photometric and spectroscopic observations. Section 5 discusses our results and mass-loss rate estimates (with and without taking into account stellar sizes). Our conclusions are in Section 6. 

\section{Overview of V444 Cyg photometric and spectroscopic studies}

\subsection{Photometric studies and $\dot{P}$ estimates for V444 Cyg}

Optical light curves of V444 Cyg and their interpretation have been published many times. First photographic data obtained in 1916-1917 were published by \cite{Gaposchkin1941}. \cite{Kron1943} obtained first photoelectric light curves ($\lambda_\mathrm{eff} = 4500$ \AA, 7200 \AA). \cite{Hiltner1949} obtained near-UV light curve of the system. A close to UBV photometry of \V\ was reported in \cite{Guseizade1965}.  
\cite{Kuhi1968} published photoelectric spectrum-scanner observations of \V. \cite{Cherepashchuk1972,Cherepashchuk1973} carried out narrow-band photoelectric observations at ($\lambda_\mathrm{eff} = 4244$, 4789, 5806, 6320, 7512 \AA). 
These observations were used to interpret light curves of \V\ and obtain the binary system parameters (see, e.g., \citealt{Cherepashchuk1975,Antokhin2001}).  \cite{Khaliullin1973} also published UBV photometry of \V.   
\cite{Cherepashchuk1984} analyzed OAO-2 observations. 
A differential V photometry was reported by   
\cite{Moffat1986} and \cite{Underhill1990}.  
\cite{Marchenko1998} analyzed lightcurves of several eclipsing systems with WR stars obtained by Hipparcos. Later studies include papers by \cite{Janiashvili2016}, \cite{Eriş2011} and \cite{Laur2017}. 

A secular increase in the orbital period of \V\ was suspected in \cite{Koch1970}, although somewhat earlier  \cite{Semeniuk1968} attempted to measure $\dot{P}$ using available data at that time. The first dynamical estimate of $\dot{P}$ was obtained by \cite{Khaliullin1974}: $\dot{P} = 0.22\pm0.04$ s~yr$^{-1}$, which was confirmed by \cite{Kornilov1979}  
using additional observations published later in \cite{Kornilov1983}. 
Later on, \cite{Khaliullin1984} found $\dot{P}=0.202\pm0.018$~s~yr$^{-1}$, while \cite{Underhill1990} measured a smaller value $\dot{P} = 0.0866\pm0.0011$.  \cite{Janiashvili2016} reported $0.202\pm0.019$ ~s~yr$^{-1}$.

\subsection{Spectroscopic studies of V444 Cyg}

The spectral binarity of HD 193576 = V444 Cyg was first reported by \cite{Wilson1939}. Two spectrograms obtained in consecutive nights revealed a significant change in positions of spectral lines, different for emissions and absorptions (+425 \kms\ and -170 \kms, respectively). \cite{Wilson1940} analyzed 40 spectrograms and calculated elements of circular orbits of the components. 
\cite{Keeping1947} measured radial velocities of stars in \V\ using 55 spectrograms. The obtained orbital solution allows a weak orbital eccentricity ($e=0.026\pm0.014$), which, however, could be spurious \citep{Lucy1971}.

The earliest detailed spectroscopic study of \V\ was reported by \cite{Münch1950}. 
Variations of the radial velocity and intensities  of different spectral features of the eclipsing binary system HD 193576 were analyzed using 93 spectra obtained in 1949. The radial velocity curve of the O component measured using hydrogen Balmer absorptions H8, H9 and H10 enabled the determination of the $\gamma$-velocity and semi-amplitude: $\gamma=+10$ \kms, $K=120$ \kms. 
For the WR component, the radial velocity amplitude was found to be $K=50$ \kms, but the $\gamma$-velocity of lines \ion{N}{V} was shifted redwards by 60 \kms\ relative to the systematic velocity of H lines, while the $\gamma$-velocity of $\lambda$4058 \ion{N}{IV} showed a blueshift of 50 \kms. The author explained the shift of the emission lines by absorption in the WR shell. 

\cite{Ganesh1967} reported 28 spectroscopic measurements of \V\ and derived orbital solutions from them. 

\cite{Underhill1988} performed spectroscopy of \V\ in a yellow-green spectral band to derive some orbital elements. Using the \ion{He}{II} $\lambda$5411 line, they found $\gamma_{\mathrm{WR}} = +100\pm8$ \kms,  $K_{\mathrm{WR}} = 370\pm12$ \kms; the N IV $\lambda$5203 yielded: $\gamma_{\mathrm{WR}} = -40\pm12$ \kms, $K_{\mathrm{WR}} = 337\pm18$ \kms; for the O-star $\gamma_O = +22\pm5$ \kms, $K_O = 112\pm8$ \kms. From \ion{He}{II} and \ion{O}{III} lines the masses of the WR and O component were found to be  11.3 and 37.5 $M_{\sun}$, respectively, while  the lines \ion{N}{IV} and \ion{O}{III} gave 9.8 and 29.6 $M_{\sun}$, respectively. 

\cite{Acker1989} reported data on spectroscopic variations of \V\ from observations carried out between 1979 and 1982 with a mean resolution of 2 \AA\ in the red part of the spectrum. It was found that the ratio of the radial velocity curves of emission (the WN5 star) and absorption (the O star) lines is about factor two yielding a binary mass ratio $q = M(WN5)/M(O6)$ of 0.4. The observed radial velocity amplitudes are higher than measured by \cite{Münch1950}. 

Numerous optical spectra of \V\ with a high signal-to-noise ratio were obtained by S. Marchenko and his colleagues enabled a detailed analysis of the emission and absorption radial velocity curves \citep{Marchenko1994,Marchenko1997}. \cite{Marchenko1994}, in addition to presenting radial velocity curves, improved orbital parameters of the system and discussed effects of interaction visible in neutral helium lines. \cite{Marchenko1997}, based on spectral line profiles, investigated the WR stellar wind parameters and physical conditions in a wind collision region between the two components.

The latest  \V\ radial velocity measurements are presented in \cite{Dsilva2022}.  

\begin{figure*}
  \centering
  \includegraphics[scale=0.53]{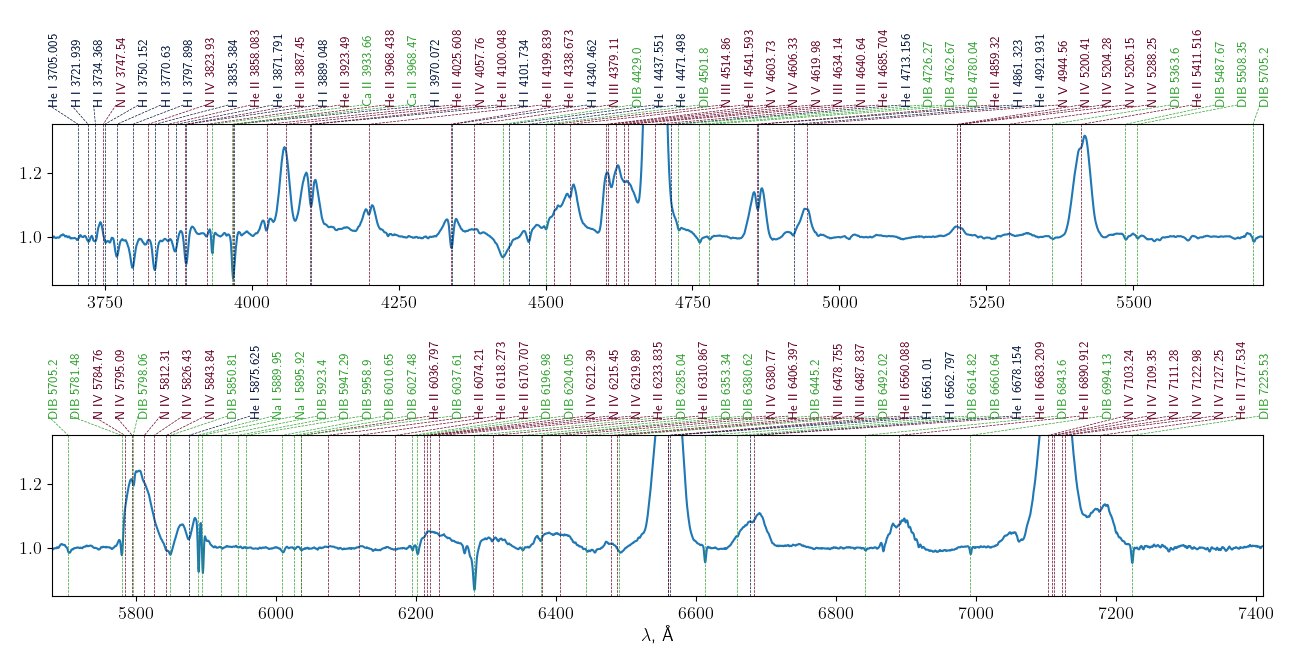}
  \caption{Average spectrum of \V. WR emission lines are noted in dark red, O-star absorptions are in dark blue, interstellar features are in green.}
 \label{mean}
\end{figure*}

\section{Observations and data analysis}

Our spectroscopic observations of \V\ were performed 
for 20 nights from October 2022 till 
August 2023 on the 2.5-m telescope of the Caucasian Mountain Observatory of Sternberg Astronomical Institute (CMO SAI) with the low-resolution TDS spectrograph \citep{TDS}. 

In each of the 20 nights, the final spectrum was obtained by averaging three consecutive 5-min exposures. To avoid pileup, the star was shifted along the slit. This long exposure is necessary to obtain a reliable sky line spectrum, which we use to additional correction of the wavelength calibration. After this correction, the accuracy of the calibration is about 3 \kms\ in the red channel (5700-7400 \AA\ ) and worsens in the blue channel (3600-5700 \AA\ ) due to the lack of the telluric lines, from 5 \kms\ in the range of 4700-5700 \AA\ to 20 \kms\ at  3600 \AA.

Fig. \ref{mean} shows the average spectrum of \V\ with identified spectral features.

\begin{figure}
  \centering
  \includegraphics[scale=0.37]{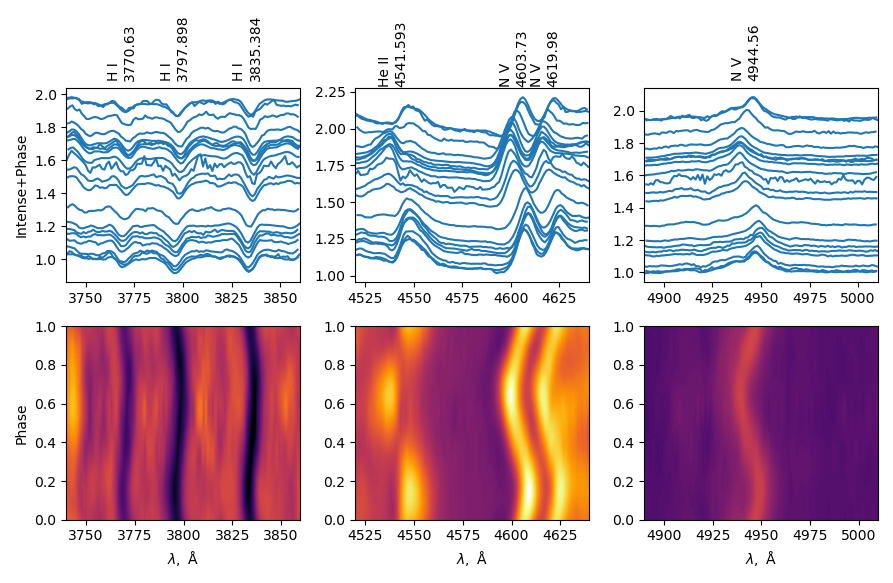}
  \caption{
  Fragments of the obtained spectra near the selected lines (top) and dynamic spectra constructed from these fragments (bottom).}
 \label{dynsp}
\end{figure}

We measured radial velocities for selected lines by fitting the profiles with a Gaussian. We selected three ultraviolet Balmer hydrogen absorption lines and three emission lines of the \ion{N}{V} ion.  
Fig. \ref{dynsp} shows spectral fragments with a resolution of $R\sim 1500$ around used lines and interpolated dynamic spectra.
We calculated semi-amplitudes and $\gamma$-velocities for \ion{N}{V} $\lambda$4603+4619 ($K^{NV}_{4603-19} = 306\pm4$ \kms, $\gamma^{NV}_{4603-19} = 51\pm3$ \kms), \ion{N}{V} $\lambda$4945 ($K^{NV}_{4945} = 309\pm5$ \kms, $\gamma^{NV}_{4945} = -34\pm4$ \kms) and \ion{H}{I} $\lambda$3770+3798+3835 ($K^{HI} = 124\pm11$ \kms, $\gamma^{HI} = 92\pm10$ \kms).

\begin{figure}
  \centering
  \includegraphics[scale=0.48]{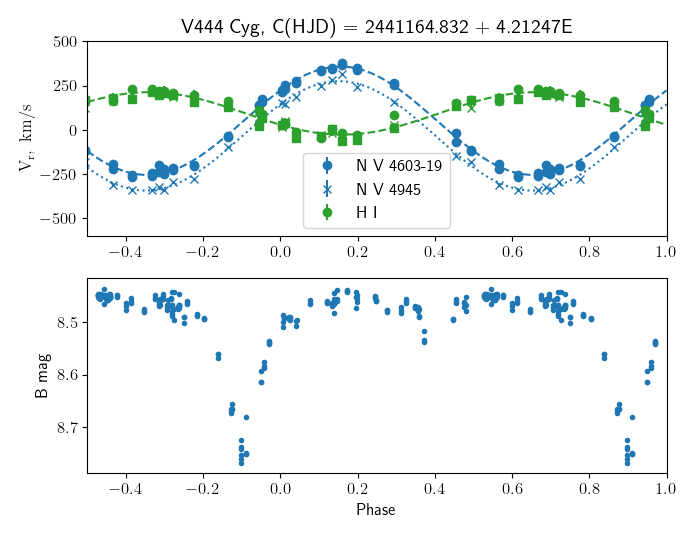}
  \caption{Radial velocity curves of \V\ in lines \ion{N}{V}, \ion{H}{I} (upper) and B lightcurve (lower) from our CMO observations. }  
 \label{RVCs}
\end{figure}

Table \ref{vels} lists the radial velocity measurements in selected lines. Fig. \ref{RVCs} 
(upper panel)
shows the radial velocity curves for lines \ion{H}{I} and \ion{N}{V}. 

\begin{table}
    \centering
    \scriptsize
    \caption{Radial velocities from the \V\ spectra (\kms).}
    \begin{tabular}{ccccccc}
\hline
HJD         &  V$^{\ion{H}{I}}_{3770}$  &  V$^{\ion{H}{I}}_{3798}$  &  V$^{\ion{H}{I}}_{3835}$  &  V$^{\ion{N}{V}}_{4603}$ &  V$^{\ion{N}{V}}_{4619}$  &  V$^{\ion{N}{V}}_{4945}$  \\
\hline
2459855.32548 &  113$\pm$15  &   20$\pm$22  &   72$\pm$20  &  141$\pm$6  &   141$\pm$6  &    65$\pm$7   \\
2459883.22391 &  164$\pm$21  &  178$\pm$28  &  165$\pm$26  & -222$\pm$7  &  -193$\pm$7  &  -312$\pm$13  \\
2459885.21906 &  -13$\pm$19  &  -44$\pm$15  &  -20$\pm$13  &  275$\pm$6  &   263$\pm$6  &   185$\pm$6   \\
2459896.36757 &  218$\pm$20  &  196$\pm$20  &  205$\pm$14  & -239$\pm$6  &  -200$\pm$7  &  -321$\pm$6   \\
2459906.15989 &   35$\pm$18  &   41$\pm$21  &   50$\pm$16  &  228$\pm$5  &   253$\pm$7  &   145$\pm$8   \\
2459913.26741 &  220$\pm$28  &  207$\pm$26  &  201$\pm$13  & -249$\pm$6  &  -220$\pm$10 &  -340$\pm$5   \\
2459929.25589 &  170$\pm$17  &  168$\pm$21  &  125$\pm$15  & -117$\pm$7  &  -121$\pm$6  &  -183$\pm$8   \\
2459930.21664 &  206$\pm$15  &  199$\pm$14  &  188$\pm$13  & -226$\pm$5  &  -215$\pm$6  &  -296$\pm$6   \\
2459931.19100 &   91$\pm$15  &   64$\pm$18  &   81$\pm$13  &  155$\pm$6  &   176$\pm$6  &    71$\pm$6   \\
2459932.22060 &  -30$\pm$17  &  -59$\pm$19  &  -34$\pm$13  &  335$\pm$6  &   349$\pm$6  &   240$\pm$7   \\
2459934.20018 &  233$\pm$17  &  213$\pm$14  &  211$\pm$13  & -263$\pm$5  &  -246$\pm$6  &  -342$\pm$5   \\
2459950.14895 &  149$\pm$15  &  136$\pm$15  &  147$\pm$13  &  -70$\pm$5  &   -19$\pm$6  &  -148$\pm$6   \\
2459964.14620 &  196$\pm$27  &  157$\pm$20  &  163$\pm$20  & -204$\pm$6  &  -198$\pm$8  &  -279$\pm$6   \\
2459965.65385 &    6$\pm$17  &    2$\pm$26  &  -20$\pm$14  &  348$\pm$6  &   345$\pm$6  &   279$\pm$7   \\
2460029.51541 &   84$\pm$19  &   11$\pm$19  &   25$\pm$14  &  251$\pm$6  &   267$\pm$6  &   158$\pm$6   \\
2460032.50278 &   27$\pm$19  &   26$\pm$14  &   13$\pm$13  &  222$\pm$6  &   212$\pm$7  &   152$\pm$6   \\
2460061.40325 &  164$\pm$20  &  131$\pm$15  &  148$\pm$18  &  -36$\pm$6  &   -40$\pm$10 &   -98$\pm$7   \\
2460138.47417 &  -20$\pm$15  &  -64$\pm$18  &  -65$\pm$13  &  376$\pm$5  &   368$\pm$9  &   315$\pm$6   \\
2460142.44906 &  -46$\pm$19  &  -33$\pm$19  &  -31$\pm$14  &  340$\pm$5  &   333$\pm$6  &   247$\pm$9   \\
2460178.29902 &  228$\pm$15  &  175$\pm$14  &  205$\pm$13  & -268$\pm$5  &  -254$\pm$6  &  -338$\pm$6   \\
    \end{tabular}
    \label{vels}
\end{table}

In addition, in winter 2022-2023 and summer 2023, broadband multicolor photometric observations of \V\ were carried out on the 60-cm RC-600 reflector of CMO SAI \citep{RC600}. The obtained 
light curve in B filter is
presented in Fig. \ref{RVCs} 
(lower panel)
Unfortunately, we were unable to observe the secondary minimum, and so far we have not used these data to calculate the system parameters. We do use the B observations to compare with extant light curves of \V\ (see the next Section). 

\section{Search for evolutionary orbital period changes in V444 Cyg}

The orbital period change rate $\dot{P}$ can be most reliably calculated by a modified Hertzsprung method from comparison of the light curves at the (usually primary) minimum. In this method, one of the light curve is taken as a template which is overlaid with other light curves to search for phase shifts. This results in the time dependence of the phase shifts between the curves enabling the estimation of the binary system parameters change. For example, a quadratic fit  to the obtained dependence $O-C = A\cdot E^2 + B\cdot E + C$ makes it possible to estimate the orbital period change rate $\dot{P} = 2\cdot A$.  

\begin{figure}
  \centering
  \includegraphics[scale=0.5]{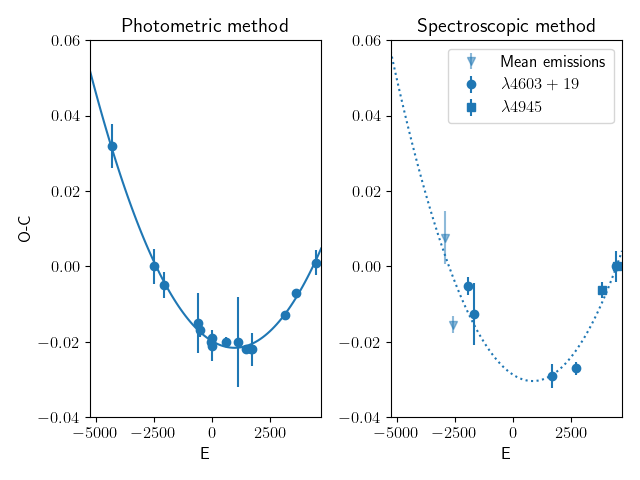}
  \caption{O-C plots (in fractions of the orbital period) calculated from photometric data (the left panel) and spectroscopic radial velocity curves (the right panel) from Table \ref{OC-tab}}.
 \label{OC-pic}
\end{figure}

\V\ is the first WR+OB system in which a secular change in the orbital period was discovered by this method. Here we repeat such a calculation using the new data. Table \ref{OC-tab} provides information on the light curves used and calculated residuals. The O-C diagram for the photometric data is shown in the left panel of Fig. \ref{OC-pic}. 

At the same time, as shown in \citepalias{Shaposhnikov2023}, a comparison of radial velocity curves can also be used to estimate $\dot{P}$. We should specially note that the radial velocity curves should be selected with caution because, at first, radial velocity curves for different ions in WR spectra can demonstrate different $\gamma$-velocities and, at second, many radial velocity curves are affected by  spectral line profile distortions. For WN stars, the best choice is \ion{N}{V} lines due to a high ionization potential.
This guarantees that the observed lines are formed at the very base of the accelerated stellar wind.

Table \ref{OC-tab} summarizes the archival data used by us and our new observations. The calculated O-C diagram is presented in the right panel of Fig. \ref{OC-pic}. 

\begin{table}
    \centering
    \scriptsize
    \caption{Photometric and spectroscopic data of \V\ used to calculate O-C (in fractions of the orbital period).}
    \begin{tabular}{cccc}
\hline
Mean HJD & Reference         & Filter                          & O-C                  \\
\hline                                                                                
2423001 & \defcitealias{Gaposchkin1941}{Gaposchkin (1941)}      \citetalias{Gaposchkin1941}     & IPg                             &    0.032$\pm$0.006   \\
2430589 & \defcitealias{Kron1943}{Kron (1943)}            \citetalias{Kron1943}           & $\lambda_\mathrm{eff} = 4500$\AA   &   0.0000$\pm$0.0005  \\
2432448 & \defcitealias{Hiltner1949}{Hiltner (1949)}         \citetalias{Hiltner1949}        & $\lambda_\mathrm{eff} = 3550$\AA   &   -0.005$\pm$0.003   \\
2438638 & \defcitealias{Guseizade1965}{Guseizade (1965)}       \citetalias{Guseizade1965}      & B                               &   -0.015$\pm$0.008   \\
2438969 & \defcitealias{Kuhi1968}{Kuhi (1968)}           \citetalias{Kuhi1968}           & $\lambda_\mathrm{eff} = 4786$\AA   & -0.0170$\pm$0.0018 \\
2441201 & \defcitealias{Khaliullin1973}{Khaliullin (1973)}      \citetalias{Khaliullin1973}     & B                               &  -0.0190$\pm$0.0020  \\
2441267 & \defcitealias{Cherepashchuk1984}{Cherepashchuk (1984)}   \citetalias{Cherepashchuk1984}  & $\lambda_\mathrm{eff} = 4250$\AA & -0.021$\pm$0.004  \\
2443719 & \defcitealias{Kornilov1983}{Kornilov (1983)}        \citetalias{Kornilov1983}       & B & -0.0200$\pm$0.0012 \\
2445245 & \defcitealias{Cherepashchuk1972}{Cherepashchuk (1972)}   \citetalias{Cherepashchuk1972}  & $\lambda_\mathrm{eff} = 4244$\AA   &  -0.0200$\pm$0.0010  \\
2445887 & \defcitealias{Moffat1986}{Moffat (1986)}          \citetalias{Moffat1986}         & B & -0.020$\pm$0.012 \\
2447408 & \defcitealias{Underhill1990}{Underhill (1990)}       \citetalias{Underhill1990}      & V                               &  -0.0220$\pm$0.0005  \\
2448181 & \defcitealias{Janiashvili2016}{Janiashvili (2016)}     \citetalias{Janiashvili2016}    & V &-0.0220$\pm$0.0010 \\ 
2448355 & \defcitealias{Marchenko1998}{Marchenko (1998)}       \citetalias{Marchenko1998}      & Hp & -0.022$\pm$0.004 \\
2454344 & \defcitealias{Eriş2011}{Eriş (2011)}            \citetalias{Eriş2011}           & B                               &  -0.0130$\pm$0.0020  \\
2456362 & \defcitealias{Laur2017}{Laur (2017)}            \citetalias{Laur2017}           & B                               &  -0.0070$\pm$0.0005  \\
2460021 & Our data                  & B                               &    0.001$\pm$0.004   \\
\hline
Mean HJD    & Reference & Lines & O-C         \\
\hline
2428773  & \defcitealias{Wilson1940}{Wilson (1940)}            \citetalias{Wilson1940}         & Mean emiss.       &   0.008$\pm$0.007  \\
2430307  & \defcitealias{Keeping1947}{Keeping (1947)}         \citetalias{Keeping1947}        & Mean emiss.       &  -0.015$\pm$0.002  \\ 
2433125  & \defcitealias{Münch1950} {Münch (1950)}             \citetalias{Münch1950}          & \ion{N}{V}, $\lambda$4603+19 &  -0.005$\pm$0.002  \\
2434173  & \defcitealias{Ganesh1967}{Ganesh (1967)}            \citetalias{Ganesh1967}         & \ion{N}{V}, $\lambda$4603 &  -0.013$\pm$0.008  \\
2448327  & \defcitealias{Marchenko1994}{Marchenko (1994)}      \citetalias{Marchenko1994}      & \ion{N}{V}, $\lambda$4603 &  -0.029$\pm$0.003  \\
2452729  & \defcitealias{Hirv2006}{Hirv (2006)}              \citetalias{Hirv2006}           & \ion{N}{V}, $\lambda$4603 & -0.0271$\pm$0.0017 \\
2457422  & \defcitealias{Dsilva2022}{Dsilva (2022)}            \citetalias{Dsilva2022}         & \ion{N}{V}, $\lambda$4945 & -0.0063$\pm$0.0021 \\
2460022  & Our data                  & \ion{N}{V}, $\lambda$4603+19, 4945          & 0.0000$\pm$0.0005  \\

\end{tabular}
    \label{OC-tab}
\end{table}

\begin{table}
    \centering
    \caption{Parabolic fit coefficients in Fig.\ref{OC-pic} and $\dot{P}$ estimates.}
    \begin{tabular}{ccccc}
\hline
Data         & A, $10^{-9}$ & B, $10^{-6}$    & C, $10^{-2}$   & $\dot{P}$, \syr   \\
\hline
Phot. & 1.89$\pm$0.05 & -3.69$\pm$0.14 & -1.97$\pm$0.04 & 0.119$\pm$0.003 \\
Spec. & 2.33$\pm$0.51 &  -4.0$\pm$1.1  &  -2.9$\pm$0.5  &  0.147$\pm$0.032  \\
\end{tabular}
    \label{parab_coef}
\end{table}

The O-C diagrams calculated by both photometric and spectroscopic methods demonstrate a remarkable similarity. Table \ref{parab_coef} presents the coefficients of the parabolic fits for two cases and the derived values of the orbital period change rate $\dot{P}$. Clearly, both methods yield consistent results, however, errors for the spectroscopic method are significantly higher. Therefore, the final estimate of $\dot{P} = 0.127\pm0.004$ \syr\ in \V\ can be taken from the photometric analysis. 

\section{Discussion. Estimates of $\dot{M}$}

The spectroscopic orbit elements of \V\ calculated by us are in good agreement with earlier results. Our estimate of  $\dot{P}$ is also consistent with previous studies. Given the derived  $\dot{P}$, the orbital period of \V\ at the time of writing (MJD = 60158) is $P \simeq 4.21250^d$. From the actual value of the orbital period and the semi-amplitudes of radial velocity curves (based on \ion{N}{V} $\lambda$4603-19 and \ion{H}{I} lines), we can calculate masses of the components and the size of the orbit (to within the usual $\sin{i}$ uncertainty): 
$M_{\mathrm{O}}\sin^3{i} = 24.7\pm0.3~M_{\sun}$, 
$M_{\mathrm{WR}}\sin^3{i} = 10.0\pm0.9~M_{\sun}$, 
$a_{\mathrm{O}}\sin{i} = 10.3\pm0.9~R_{\sun}$,
$a_{\mathrm{WR}}\sin{i} = 25.5\pm0.3~R_{\sun}$; 
$q = M_{\mathrm{WR}}/M_{\mathrm{O}} = 0.41\pm0.04$.
By adopting $i=78^\circ$ from the light curve solution in \citealt{Antokhin2001}), we get
$M_{\mathrm{O}} \approx 26.4~M_{\sun}$, $M_{\mathrm{WR}} \approx 10.7~M_{\sun}$, 
$a_{\mathrm{O}} \approx 10.6~R_{\sun}$, $a_{\mathrm{WR}} \approx 26.1~R_{\sun}$. 

These parameters enable us to estimate the mass-loss rate from the WR star in the simplest model. This approximation assumes a spherically symmetric stellar wind, point-like stars and ignores stellar sizes and mass transfer between the components. In this case $\dot{M}_{\mathrm{WN5}} = -\frac{1}{2} \frac{\dot{P}}{P} (M_{\mathrm{WN5}}+M_{\mathrm{O6}}) = -(5.8\pm 0.4)\cdot 10^{-6}~M_{\sun}\mbox{ yr}^{-1}.$

However, as shown in \citetalias{Shaposhnikov2023}, we can make more accurate estimates. In close binaries, finite stellar sizes of synchronously rotating components and possible mass exchange between them affect the mass-loss estimates. In this case (see formula [A7] in the Appendix to \citetalias{Shaposhnikov2023})
\beqa{a:dPP}
&\displaystyle\frac{\dot{P}}{P}=-\frac{\dot{M_1}}{M_1}\left\{3+3\frac{x-\beta}{q}-\frac{\alpha+x}{1+q}\right.\\
&\displaystyle\left.-3\frac{1+q}{q}\left(
\alpha\left[\myfrac{q}{1+q}^2+\frac{2}{3}\myfrac{R_1}{a}^2\right]+
x\left[\myfrac{1}{1+q}^2+\frac{2}{3}\myfrac{R_2}{a}^2\right]
\right)
\right\}\nonumber
\eeqa
Here $M_1 = M_{\mathrm{O6}}$, $M_2 = M_{\mathrm{WN5}}$ (index "1"\ marks a more massive component), $q = M_2/M_1$ ($q<1$), $\alpha$ is the fraction of the total mass-loss rate $\dot{M_1}$ due to stellar wind, $\beta$ is the fraction of the total mass-loss rate $\dot{M_1}$ due to the mass exchange between the components ($\dot{M_1}=\dot{M_{1,w}}+\dot{M_{1,t}}=\alpha \dot{M_1}+\beta \dot{M_1},~\alpha+\beta = 1$), $x$ is the coefficient characterizing the stellar wind mass-loss rate from the less massive component ($\dot{M_2}=\dot{M_{2,w}}-\dot{M_{1,t}}=x \dot{M_1}-\beta \dot{M_1}$).

In case of \V\ we can neglect the mass exchange between the components as the system is wider than CQ Cep and significantly positive $\dot{P}$ ($\alpha=1$, $\beta=0$). The stellar radii in \V\ can be taken as $R_{\mathrm{O6}} = 8.5~R_{\sun}$, $R_{\mathrm{WN5}} = 3~R_{\sun}$ (after \citealt{Antokhin2001}). The account of only WR radius gives a small correction to the simplest model estimate of $\dot{M_{\mathrm{WN5}}}$ providing the observed value $\dot{P} = 0.119\pm0.003$ \syr: $\dot{M}_{\mathrm{WN5}} = -(6.1\pm 0.4)\cdot 10^{-6}~M_{\sun}\mbox{ yr}^{-1}$.

If we 'switch off' the WR wind and leave only a spherically symmetric wind from the O-star with radius $R_{\mathrm{O6}}$, the corresponding mass-loss rate is $\dot{M_{\mathrm{O6}}} = -(8.0\pm 0.4)\cdot 10^{-6}~M_{\sun}\mbox{ yr}^{-1}$.

\begin{figure}
  \centering
  \includegraphics[scale=0.52]{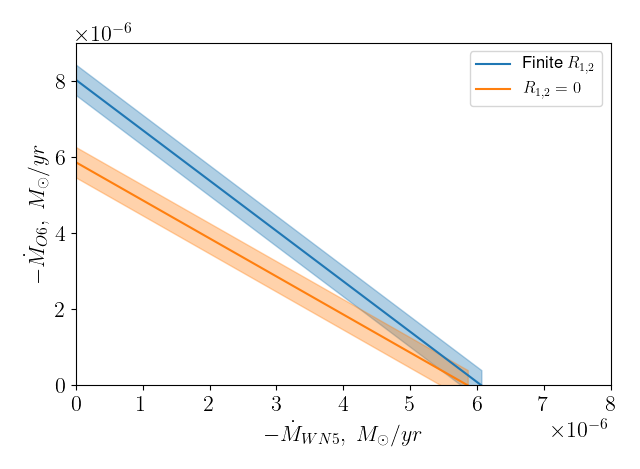}
  \caption{The relation between mass-loss rates $\dot{M}_{\mathrm{WN5}}$ and $\dot{M}_{\mathrm{O6}}$ that provides the observed value of the secular orbital period increase in \V. The model with account for finite stellar radii is in blue; the model with point-like stars is in orange.}
 \label{mdot}
\end{figure}

Fig. \ref{mdot} illustrates possible intermediate cases providing the required value of $\dot{P}$  for the calculated mass range and adopted radii of the components in \V. For $x = \dot{M}_{\mathrm{WN5}}/\dot{M}_{\mathrm{O6}} = 10$ the solution taking into account the component radii yields  $\dot{M}_{\mathrm{O6}} = -(6.0\pm 0.4)\cdot 10^{-7}~M_{\sun}\mbox{ yr}^{-1}$, $\dot{M}_{\mathrm{WN5}} = -(6.0\pm 0.4)\cdot 10^{-6}~M_{\sun}\mbox{ yr}^{-1}$.

\section{Conclusion}

We have performed new low-resolution spectral and photometric observations of WR+OB binary \V\ on 2.5-m and 60-cm RC-600 telescopes of CMO SAI. Using the extant and new  data, from photometric and spectral method we have determined the rate of a secular increase of the orbital period of \V\ $\dot{P}_{\mathrm{ph}}=0.119\pm0.003$ \syr\ and $\dot{P}_{\mathrm{sp}}=0.147\pm0.032$ \syr, respectively. The value of $\dot{P}_{\mathrm{sp}}$ derived from the analysis of radial velocity curves is in good agreement with the photometric estimate obtained from analysis of light curves of \V\ by a modified Hertzsprung method.  

The obtained  $\dot{P}$ enabled us to estimate the mass-loss rate from the WN5 star with account for the O-star mass-loss rate: $\dot{M}_{\mathrm{WN5}}=(6.0\pm0.4)\cdot10^{-6}~M_{\sun}$~yr$^{-1}$. This estimate is in good agreement with previous estimates of $\dot{M}_{\mathrm{WN5}}$ derived from photometrical data and is consistent with  $\dot{M}_{\mathrm{WN5}}=(6-7.5)\cdot10^{-6}~M_{\sun}$~yr$^{-1}$ inferred from the analysis of variable optical linear polarization of \V\ \citep{Robert1990,St-Louis1993}. We stress that estimates of $\dot{M}_{\mathrm{WR}}$ derived from the binary orbital period change and the variable linear polarization analysis are not subjected to stellar wind clumping effects and thus are more reliable. Estimates of $\dot{M}_{\mathrm{WR}}$ found from the analysis of radio and IR fluxes from WR stars are overestimated by a factor of 3-5 due to the WR wind clumping. Thus, we can conclude that the mass-loss rate from a 10.5 $M_{\sun}$ WN5 star is $(6.0\pm0.4)\cdot10^{-6}~M_{\sun}$~yr$^{-1}$.

It is interesting to estimate mass-loss rates from WR stars of different types, masses, spectral classes and chemical compositions. This could be done from the analysis of the orbital period changes of many known WR+OB binaries. Unfortunately, the number of eclipsing systems among them is small, and application of the photometric method is restricted. 

Based on the analysis of three eclipsing WR+OB binaries CQ Cep, CX Cep and V444 Cyg (\citetalias{Shaposhnikov2023} and the present work), we have shown that the rate of secular change of the orbital period in a WR+OB system obtained by the spectral method is in good agreement with estimates obtained by the photometric method. This suggests that the spectral method is reliable and prospective to determine secular changes in orbital periods of many (several dozens) spectral double WR+OB systems harboring different types of WR stars. Importantly, the initial epochs of spectral observations of many WR+OB systems date back to 1950-1960. This provides a large epoch difference compared to the present observations, which is favorable to search for secular orbital period changes in these systems. Therefore, further spectral observations of numerous different WR+OB binary systems are very prospective and useful. 

\section*{Acknowledgements}

The authors thank the referee, Prof. Petr Harmanec, for useful notes, which helped us to improve the paper.
The work of ISh, ACh and AD (observations and data analysis) is supported by the Russian Science Foundation through grant 23-12-00092. 
The authors acknowledge Drs. N.P. Ikonnikova and M.A. Burlak for carrying out photometric observations. 

\section*{Data Availability} 

Data of spectroscopic and photometric observations are available on reasonable request from the authors.

\label{lastpage}

\begin{thebibliography}{}
\makeatletter
\relax
\def\mn@urlcharsother{\let\do\@makeother \do\$\do\&\do\#\do\^\do\_\do\%\do\~}
\def\mn@doi{\begingroup\mn@urlcharsother \@ifnextchar [ {\mn@doi@}
  {\mn@doi@[]}}
\def\mn@doi@[#1]#2{\def\@tempa{#1}\ifx\@tempa\@empty \href
  {http://dx.doi.org/#2} {doi:#2}\else \href {http://dx.doi.org/#2} {#1}\fi
  \endgroup}
\def\mn@eprint#1#2{\mn@eprint@#1:#2::\@nil}
\def\mn@eprint@arXiv#1{\href {http://arxiv.org/abs/#1} {{\tt arXiv:#1}}}
\def\mn@eprint@dblp#1{\href {http://dblp.uni-trier.de/rec/bibtex/#1.xml}
  {dblp:#1}}
\def\mn@eprint@#1:#2:#3:#4\@nil{\def\@tempa {#1}\def\@tempb {#2}\def\@tempc
  {#3}\ifx \@tempc \@empty \let \@tempc \@tempb \let \@tempb \@tempa \fi \ifx
  \@tempb \@empty \def\@tempb {arXiv}\fi \@ifundefined
  {mn@eprint@\@tempb}{\@tempb:\@tempc}{\expandafter \expandafter \csname
  mn@eprint@\@tempb\endcsname \expandafter{\@tempc}}}
\def\pasp{PASP}


\bibitem[\protect\citeauthoryear{Acker, Prevot, \& Prevot}{Acker et~al.}{1989}]{Acker1989}
 Acker, A., Prevot, M. L., \& Prevot, L., 1989, \aap, 226(1), 137

\bibitem[\protect\citeauthoryear{Antokhin \& Cherepashchuk}{Antokhin \&
  Cherepashchuk}{2001}]{Antokhin2001}
 Antokhin, I. I., \& Cherepashchuk, A. M., 2001, \mn@doi [Astron. Rep.] {10.1134/1.1369800}, 45, 371

\bibitem[\protect\citeauthoryear{Berdnikov et al.}{Berdnikov et al.}{2020}]{RC600}
 Berdnikov, L. N.,  Belinskii, A. A., Shatskii, N. I., Burlak, M. A., Ikonnikova, N. P., Mishin, E. O.,  Cheryasov, D. V., Zhuiko, S. V., 2020,  \mn@doi
 [Astronomy Reports] {10.1134/S1063772920040010}, 64, 310

\bibitem[\protect\citeauthoryear{Cherepashchuk}{Cherepashchuk}{1975}]{Cherepashchuk1975}
 Cherepashchuk, A. M., 1975, SvA, 19, 47

\bibitem[\protect\citeauthoryear{Cherepashchuk \& Khaliullin}{Cherepashchuk \& Khaliullin}{1972}]{Cherepashchuk1972}
 Cherepashchuk, A. M., \& Khaliullin, K. F., 1972, Peremennye Zvezdy, 18, 321

\bibitem[\protect\citeauthoryear{Cherepashchuk \& Khaliullin}{Cherepashchuk \& Khaliullin}{1973}]{Cherepashchuk1973}
 Cherepashchuk, A. M., \& Khaliullin, K. F., 1973, SvA, 17, 330

\bibitem[\protect\citeauthoryear{Cherepashchuk, Eaton \& Khaliullin}{Cherepashchuk et~al.}{1984}]{Cherepashchuk1984}
 Cherepashchuk, A. M., Eaton, J.A. \& Khaliullin, K. F., 1984, \mn@doi [\apj] {10.1086/162156}, 281, 774

\bibitem[\protect\citeauthoryear{Dsilva et~al.}{Dsilva et~al.}{2022}]{Dsilva2022}
 Dsilva, K., Shenar, T., Sana, H., Marchant, P., 2022, \mn@doi [\aap] {10.1051/0004-6361/202142729}, 664, A93

\bibitem[\protect\citeauthoryear{Eriş \& Ekmekçi}{Eriş \& Ekmekçi}{2011}]{Eriş2011}
 Eriş, F. Z., \& F. Ekmekçi, 2011, \mn@doi [Astronomische Nachrichten] {10.1002/asna.201011564}, 332(6), 616

\bibitem[\protect\citeauthoryear{Ganesh, Bappu, \& Natarajan}{Ganesh et~al.}{1967}]{Ganesh1967}
 Ganesh, K. S., Bappu, M. K. V., \& Natarajan, V., 1967, Kodaikanal Obs. Bull., Ser. A, 184, 83

\bibitem[\protect\citeauthoryear{Gaposchkin}{Gaposchkin}{1941}]{Gaposchkin1941}
 Gaposchkin, S., 1941, \mn@doi [\aap] {10.1086/144255}, 93, 202, 93, 202
 
\bibitem[\protect\citeauthoryear{Guseizade}{Guseizade}{1965}]{Guseizade1965}
 Guseizade, A. A., 1965, Peremennye Zvezdy, 15, 555

\bibitem[\protect\citeauthoryear{Hiltner}{Hiltner}{1949}]{Hiltner1949}
 Hiltner, W. A., 1949, \mn@doi [\aap] {10.1086/145184}, 110, 95
 
\bibitem[\protect\citeauthoryear{Hirv, Annuk, Eenmäe, Liimets, Pelt, Puss, \& Tempel}{Hirv et~al.}{2006}]{Hirv2006}
 Hirv, A., Annuk, K., Eenmäe, T., Liimets, T., Pelt, J., Puss, A., \& Tempel, M., 2006, Baltic Astronomy, 15, 405-412.

\bibitem[\protect\citeauthoryear{Janiashvili \& Urushadze}{Janiashvili \& Urushadze}{2016}]{Janiashvili2016}
 Janiashvili, E. B., \& Urushadze, T. V., 2016, Samtskhe-Javakheti State University Press, 1, 17

\bibitem[\protect\citeauthoryear{Keeping}{Keeping}{1947}]{Keeping1947}
 Keeping, E. S., 1947, Publications of the Dominion Astrophysical Observatory Victoria, 7, 349
 
\bibitem[\protect\citeauthoryear{Khaliullin}{Khaliullin}{1973}]{Khaliullin1973}
 Khaliullin, K. F., 1973, Peremennye Zvezdy, 19, 73
 
\bibitem[\protect\citeauthoryear{Khaliullin}{Khaliullin}{1974}]{Khaliullin1974}
 Khaliullin, K. F., 1974, SvA, 18, 229.

\bibitem[\protect\citeauthoryear{Khaliullin, Khaliullina, \& Cherepashchuk}{Khaliullin et~al.}{1984}]{Khaliullin1984}
 Khaliullin, K. F., Khaliullina, A. I., \& Cherepashchuk, A. M., 1984, SvAL, 10, 250.

\bibitem[\protect\citeauthoryear{Koch}{Koch}{1970}]{Koch1970}
 Koch R.H., 1970, In: Mass Loss and Evolution in Close Binaries. Proc. IAU Coll. №6. Copenhagen Univ. 1970. 65.

\bibitem[\protect\citeauthoryear{Kornilov \& Cherepashchuk}{Kornilov \& Cherepashchuk}{1979}]{Kornilov1979}
 Kornilov, V. G., \& Cherepashchuk, A. M., 1979, SvAL, 5, 214.

\bibitem[\protect\citeauthoryear{Kornilov}{Kornilov}{1983}]{Kornilov1983}
 Kornilov, V. G., 1983, Peremennye Zvezdy, 21(6), 835.

\bibitem[\protect\citeauthoryear{Kron \& Gordon}{Kron \& Gordon}{1943}]{Kron1943}
 Kron, G. E., \& Gordon, K. C., 1943, \mn@doi [\apj] {10.1086/144525}, 97, 311

\bibitem[\protect\citeauthoryear{Kuhi}{Kuhi}{1968}]{Kuhi1968}
 Kuhi, L. V., 1968, \mn@doi [\apj] {10.1086/149527}, 152, 89
 
\bibitem[\protect\citeauthoryear{Lamers \& Cassinelli}{Lamers \& Cassinelli}{1999}]{Lamers1999}
 Lamers, H. J., \& Cassinelli, J. P., 1999, Introduction to stellar winds. Cambridge University Press.

\bibitem[\protect\citeauthoryear{Laur, Kolka, Eenmäe, Tuvikene, \& Leedjärv}{Laur et~al.}{2017}]{Laur2017}
 Laur, J., Kolka, I., Eenmäe, T., Tuvikene, T., \& Leedjärv, L., 2017, \mn@doi [\aap] {10.1051/0004-6361/201629395}, 598, A108

\bibitem[\protect\citeauthoryear{Lucy, \& Sweeney}{Lucy, \& Sweeney}{1971}]{Lucy1971}
Lucy, L. B., Sweeney, M. A., 1971, \mn@doi [\aap] {10.1086/111159}, 76, 544

\bibitem[\protect\citeauthoryear{Marchenko, Moffat, Eenens, Cardona, Echevarria, \& Hervieux}{Marchenko et~al.}{1997}]{Marchenko1997}
 Marchenko, S. V., Moffat, A. F. J., Eenens, P. R. J., Cardona, O., Echevarria, J., \& Hervieux, Y., 1997, \mn@doi [\apj] {10.1086/304435}, 485(2), 826

\bibitem[\protect\citeauthoryear{Marchenko, Moffat \& Koenigsberger}{Marchenko et~al.}{1994}]{Marchenko1994}
 Marchenko, S. V., Moffat, A. F. J., \& Koenigsberger, G., 1994, \mn@doi [\apj] {10.1086/173773}, 422, 810

\bibitem[\protect\citeauthoryear{Marchenko et~al.}{Marchenko et~al.}{1998}]{Marchenko1998}
 Marchenko, S. V. Moffat, A. F. J. van der Hucht, K. A. Seggewiss, W. Schrijver, H. Stenholm, B. Lundstrom, I. Setia Gunawan, D. Y. A. Sutantyo, W. van den Heuvel, E. P. J. de Cuyper, J. -P. Gomez, A. E., 1998, A\&A, 331, 1022

\bibitem[\protect\citeauthoryear{Moffat \& Shara}{Moffat \& Shara}{1986}]{Moffat1986}
 Moffat, A. F. J., Shara, M. M., 1986, \apj, 92(4), 952
 
\bibitem[\protect\citeauthoryear{Münch}{Münch}{1950}]{Münch1950}
 Münch, G., 1950, \mn@doi [\apj] {10.1086/145341}, 112, 266

 \bibitem[\protect\citeauthoryear{Potanin, et al.}{Potanin et al.}{2020}]{TDS}
 Potanin, S.A.,  Belinski, A.A., Dodin, A.V.,  Zheltoukhov, S.G., Lander, V.Yu., Postnov, K.A., et al. 2020, \mn@doi [Astronomy Letters] {10.1134/S1063773720120038}, 46, 836

\bibitem[\protect\citeauthoryear{Robert, Moffat, Bastien, St-Louis \& Drissen}{Robert et~al.}{1990}]{Robert1990}
 Robert, C., Moffat, A. F., Bastien, P., St-Louis, N., \& Drissen, L., 1990, \mn@doi [\apj] {10.1086/169051}, 359, 211

\bibitem[\protect\citeauthoryear{Semeniuk}{Semeniuk}{1968}]{Semeniuk1968}
 Semeniuk I., 1968, Acta Astronomica, 18, 313

\bibitem[\protect\citeauthoryear{Shaposhnikov, Cherepashchuk, Dodin \& Postnov}{Shaposhnikov et~al.}{2023}]{Shaposhnikov2023}
 Shaposhnikov, I., Cherepashchuk, A., Dodin, A., \& Postnov, K., 2023, \mn@doi [\mnras] {10.1093/mnras/stad1491}, 523(1), 1524 (Paper I)

\bibitem[\protect\citeauthoryear{St-Louis, Moffat, Lapointe, Efimov, Shakhovskoj, Fox, \& Piirola}{St-Louis et~al.}{1993}]{St-Louis1993}
 St-Louis, N., Moffat, A. F. J., Lapointe, L., Efimov, Y. S., Shakhovskoj, N. M., Fox, G. K., \& Piirola, V., 1993, \mn@doi [\apj] {10.1086/172751}, 410(1), 342

\bibitem[\protect\citeauthoryear{Underhill, Grieve, \& Louth}{Underhill et~al.}{1990}]{Underhill1990}
 Underhill, A. B., Grieve, G. R. \& Louth, H., 1990, \mn@doi [\pasp] {10.1086/132698}, 102(653), 749

\bibitem[\protect\citeauthoryear{Underhill, Yang, \& Hill}{Underhill et~al.}{1988}]{Underhill1988}
 Underhill, A. B., Yang, S., \& Hill, G. M., 1988, \mn@doi [\pasp] {10.1086/132313}, 100(628), 741
 
\bibitem[\protect\citeauthoryear{Wilson}{Wilson}{1939}]{Wilson1939}
 Wilson, O. C., 1939, \mn@doi [\pasp] {10.1086/125001}, 51(299), 55

\bibitem[\protect\citeauthoryear{Wilson}{Wilson}{1940}]{Wilson1940}
 Wilson, O. C., 1940, \mn@doi [\apj] {10.1086/144180}, 91, 379

\makeatother
\end{thebibliography}
\end{document}